\documentclass[12pt]{iopart}
\usepackage{graphicx}
\usepackage{epsf}

\newcommand{\be}{\begin{equation}}\newcommand{\ee}{\end{equation}}
\newcommand{\ba}{\begin{array}{l}}\newcommand{\ea}{\end{array}}
\newcommand{\baa}{\begin{eqnarray}}\newcommand{\eaa}{\end{eqnarray}}
\newcommand{\lab}[1]{\label{#1}}\newcommand{\re}[1]{(\ref{#1})}
\newcommand{\ci}[1]{\cite{#1}}

\renewcommand{\baselinestretch}{1.3}
\begin{document}

\title{Dynamics of Dirac solitons in networks}
\author{K.K. Sabirov$^{a,b}$, D.B. Babajanov$^a$, D.U. Matrasulov$^a$ and P.G. Kevrekidis$^c$}
\address{ $^a$ Turin Polytechnic University in Tashkent, 17 Niyazov Str.,
100095,  Tashkent, Uzbekistan\\
$^b$ Tashkent University of Information Technologies, 102 Amir Temur Str., 100200, Tashkent, Uzbekistan \\
$^c$ Department of Mathematics and Statistics, University of
Massachusetts, Amherst, Massachusetts 01003-4515 USA}

\ead{d.b.babajanov@gmail.com} \vspace{10pt}
\begin{indented}
\item[] August 2018
\end{indented}

\begin{abstract}
We study dynamics of Dirac solitons in prototypical networks
modeling them by the nonlinear Dirac equation on metric graphs.
Stationary soliton solutions of the nonlinear Dirac equation on
simple metric graphs are obtained. It is shown that these
 solutions  provide reflectionless vertex
transmission of the Dirac solitons under suitable conditions. The
constraints for bond nonlinearity coefficients, conjectured to
represent necessary conditions for allowing reflectionless
transmission over a Y-junction are derived. The Y-junction
considerations are also generalized to a tree network. The
analytical results are confirmed by direct numerical simulations.
\end{abstract} \pacs{}
\maketitle

\section{Introduction}

Nonlinear evolution equations on networks have attracted much
attention recently
\ci{zar2010,zar2011,adami2011,adami-eur,adami2013,Karim2013,noja,Susanto05,caputo14,Our}.
Such interest is caused by a broad variety of potential
applications of the nonlinear wave dynamics and soliton transport
 in networks, such as Bose-Einstein condensates (BECs) in branched traps,  Josephson junction networks, the DNA double helix, polymer chains, etc.

Despite the rapidly growing interest in wave dynamics on networks,
most of the studies are mainly focused on nonrelativistic wave
equations  such as the nonlinear Schr{\"o}dinger (NLS) equation
\ci{zar2010,zar2011,adami2011,adami-eur,adami2013,Karim2013,noja,Our}.
Nevertheless, there is a number of works on the
sine-Gordon (sG)  equation
in branched systems \ci{Susanto05,caputo14,Our1}. However,
relativistic wave equations such as the nonlinear Klein-Gordon and
Dirac  equations are  important in field theory and condensed
matter physics and hence exploring them on metric graphs is of
interest in its own right.
These graphs consist of a system of bonds which are connected at one or more
vertices (branching points). The connection rule is associated with the
topology of a graph. When the bonds can be assigned a length, the
graph is called a metric graph.

In this paper we address the problem of the nonlinear Dirac
equation on simple metric graphs by focusing on conservation laws
and soliton transmission at the graph vertices. Our prototypical
example will be the Y-junction. Early studies of the nonlinear
Dirac equation date back to Thirring  \ci{Thirring} and
Gross-Neveu  \ci{Gross} models of field theory. Integrability, the
nonrelativistic limit and exact solutions of the relevant models
have been considered in, e.g.,
\ci{Steeb,Barashenkov1,Fushchich,Barashenkov2,Toyama} among
others. A potential application of the nonlinear Dirac equation to
neutrino oscillations was discussed in \ci{Ng}. Recently, the
possibility of experimental realization of Dirac solitons in
Bose-Einstein condensates in honeycomb optical lattices was
discussed in~\ci{Carr09} where the nonlinear Dirac equation was
derived. Further analysis of this setting in
Refs.~\ci{Carr11,Carr150,Carr15,Carr151} has excited a rapidly
growing interest in the nonlinear Dirac equation and its soliton
solutions (see, e.g. \ci{Saxena10}-\ci{Tran}). In
\ci{Saxena10,Saxena12,Saxena14,Saxena15,Saxena16,Saxena161}, a
detailed study of soliton solutions for different types of
nonlinearity, their stability  and discussions of conservation
laws were presented. These developments have, in turn, had an
impact also on the mathematical literature where the stability of
solutions for different forms of the nonlinear Dirac equation was
explored in \ci{Pelinovsky12,Pelinovsky14,Pelinovsky16}. In
\ci{Comech10}-\ci{Comech16}, the stability of  Gross-Neveu
solitons in both 1d and 3d and for other nonlinear Dirac models an
important energy-based stability criterion was developed. The
numerical corroboration of the stability for solitary waves and
vortices in 2D nonlinear Dirac models of the Gross-Neveu type was
considered quite recently in \ci{Saxena162}.

Dirac solitons in branched systems can, in principle, be
experimentally realized in different systems of optics, but also
envisioned elsewhere (e.g. in atomic physics etc.).

A relevant such possibility consists of the  discrete waveguide
arrays of~\ci{Tran1,Tran} that can be formulated as a branched
system to be described (in the appropriate long wavelength
limit~\cite{Tran1}) by a nonlinear Dirac equation on metric
graphs. Such networks in optical systems have been proposed, e.g.,
earlier in~\cite{eugenieva}.
More exotic possibilities can be envisioned in branched
networks of honeycomb (i.e., armchair nanoribbon) optical lattices
for atomic BECs~\cite{Carr150}, although we will not focus on the latter here.

\begin{figure}[ht!]
\includegraphics[width=160mm]{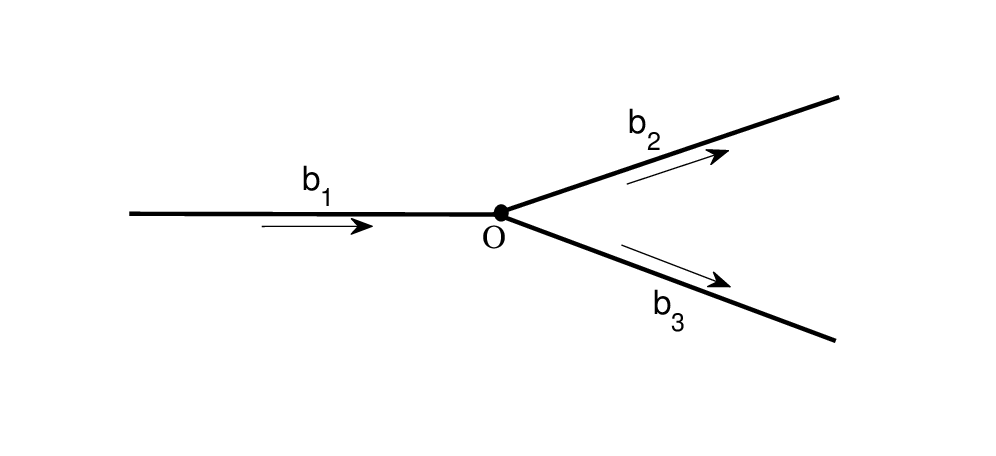} \caption{{Metric star graph
  }}
  \label{pic1}
\end{figure}

In the present work, we focus on a prototypical example of exact
solutions and transport of Dirac solitons through the network
vertices. In particular, we show that under certain constraints,
exact soliton solutions of the nonlinear Dirac equation on simple
metric graphs can be obtained. The identified soliton solutions
provide reflectionless transmission of the Dirac solitons at the
vertices. This renders possible the tuning of the transport
properties of a network in such a way that it can provide
ballistic transport of the Dirac solitons. This paper is organized
as follows: In the next section we give the formulation of the
problem on a metric star graph which includes the derivation of
the vertex boundary conditions. Section III presents soliton
solutions of the nonlinear Dirac equation on a metric star graph
for fixed and moving frames. Also, an analysis of the vertex
transmission of the Dirac solitons on the basis of the numerical
solution of the nonlinear Dirac equation is  presented. In section
IV we extend the treatment to a metric tree graph and discuss the
extension to other simple topologies. Finally, section V presents
some concluding remarks and a number of important directions
  for future work. These include a conjecture about the sufficient,
  yet not necessary nature of our vertex conditions towards ensuring
reflectionless transmission.

\section{Conservation laws and vertex boundary conditions}

The nonlinear Dirac system that we are going to explore, i.e.,
specifically the Gross-Neveu model, follows from the Lagrangian
(in the units $\hbar=c=1$) \ci{Saxena10,Saxena12,Saxena14}
\begin{eqnarray}
L = \frac{i}{2}\bar\Psi\gamma^{\mu}\partial_{\mu}\Psi
-m\bar\Psi\Psi +\frac{g^2}{2}\left(\bar\Psi\Psi\right)^2,\;\;\;
\mu=0,1,
\label{env1}
\end{eqnarray}
and describes interacting Dirac fields with $0$ and $1$
corresponding to time and coordinate variables, respectively. Here
$g$ is the nonlinearity coefficient
characterizing the strength of the nonlinear interaction. The field equations
for this Lagrangian lead to the nonlinear Dirac model of the form
\begin{equation}
\left(i\gamma^\mu\partial_\mu-m\right)\Psi
+g^2\left(\bar{\Psi}\Psi\right)\Psi=0\label{nlde1},
\end{equation}
where
\begin{equation}
\Psi(x,t)=\left(\begin{array}{cc}\phi(x,t)\\
\chi(x,t)\end{array}\right);\;\;\ \bar\Psi =\Psi^\dagger\gamma^0
=(\phi^*(x,t), -\chi^*(x,t)),\label{solit1}
\end{equation}
 and
$\gamma^0=\left(\begin{array}{cc}1&0\\0&-1\end{array}\right),\,\gamma^1=\left(\begin{array}{cc}0&1\\-1&0\end{array}\right)$.

For a metric star graph consisting of three semi-infinite bonds
(see, e.g., Fig.~\ref{pic1}), the
Lagrangian density of the nonlinear Dirac field for each
bond can be written as $L_j$ where $j$ parametrizes the bond
and $L$ for each is of the form of Eq.~(\ref{env1}).

The field equation following from this Lagrangian density can be
written as Eq.~(\ref{nlde1}) before, where the spatial coordinates
are defined as $x_1\in(-\infty,0]$ and $x_{2,3}\in[0, \infty)$,
while 0 coincides with the graph vertex.

The formulation of an  evolution set of equations on metric graphs
requires imposing vertex boundary conditions which provide
``gluing" of the graph bonds at the graph vertices. For the linear
Dirac equation on a metric graph   studied earlier by Bolte and
Harrison in \ci{Bolte}, the general vertex boundary conditions
have been derived from the self-adjointness of the Dirac operator
on a graph. In that case such conditions led to Kirchhoff rules
and continuity of the wave function at the graph vertex
\ci{Bolte}. However, for the nonlinear problem in order to derive
vertex BCs, it is arguably more natural to use suitable
conservation laws of the nonlinear flow, which give rise to
appropriate BCs. For the nonlinear Dirac system fundamental
conservation laws can be presented in terms of the momentum-energy
tensor given by \ci{Saxena10,Saxena12}
$$
T^j_{\mu\nu} =
\frac{i}{2}[\bar\Psi_j\gamma^{\mu}\partial^{\nu}\Psi_j
-\partial^{\nu}\bar\Psi_j\gamma^{\mu}\Psi_j]-g^{\mu\nu} L_j,
$$
where $g^{\mu\nu}=\left({\begin{array}{cc}1 & 0 \\ 0 &1\end{array}
}\right)$.

The energy on each bond of a star graph can be written  as
\begin{equation}
 E_j= -\int_{b_j}T^j_{00}dx  =-\int_{b_j}[\frac{i}{2}[\bar{\Psi}_j\gamma^1\partial_x\Psi_j \nonumber\\
- \partial_x\bar{\Psi}_j\gamma^1\Psi_j]-    m\bar{\Psi}_j\Psi_j+\frac{g_j^2}{2}(\bar{\Psi}_j\Psi_j)^2]dx,
\label{eq4}
\end{equation}
where integration is performed along the bond $b_j$. Here, to
derive the vertex boundary conditions (VBC) we use conservations
of charge  and energy. The total charge  for the star graph is
defined as
$$
Q=Q_1+Q_2+Q_3,
$$
where the charge for each bond is given by
\begin{equation}
Q_j=\int_{b_j}\left(|\phi_j|^2+|\chi_j|^2\right)dx.\label{eq2}
\end{equation}
The  conservation of charge (Kirchhoff rule for charge current),
$\dot{Q}=0$ yields the following vertex boundary condition: \be
\left. {\rm Re}\left[\phi_1\chi_1^*\right]\right|_{x=0}=\left.{\rm
Re}\left[\phi_2\chi_2^*\right]\right|_{x=0}+\left. {\rm
Re}\left[\phi_3\chi_3^*\right]\right|_{x=0}. \lab{vbc001} \ee Here
we used the asymptotic conditions \be \Psi_1\to0\; {\rm at}\;
x\to-\infty\; {\rm and}\; \Psi_{2,3}\to 0\; {\rm at}\; x\to\infty.
\lab{asympt1} \ee We note that the boundary conditions \re{vbc001}
can be derived also from the  conservation of the current density
given by $j =\varphi\chi^* + \chi\varphi^*$.

The energy conservation, $\dot E =0$ leads to
\baa
\left. {\rm Im}\left[\phi_1\partial_t\chi_1^*+\chi_1\partial_t\phi_1^*\right]\right|_{x=0}\nonumber\\
=\left. {\rm Im}\left[\phi_2\partial_t\chi_2^*+\chi_2\partial_t\phi_2^*\right]\right|_{x=0}+\left. {\rm Im}\left[\phi_3\partial_t\chi_3^*+\chi_3\partial_t\phi_3^*\right]\right|_{x=0}.\nonumber\\
\lab{vbc002} \eaa

As per the above analysis, the VBC  for  NLDE on metric graphs
have been obtained from the energy and charge conservation.
However, the VBC given by Eqs.~\re{vbc001} can be fulfilled, if
the following linear relations at the vertices are imposed (see
Appendix for details):
\begin{eqnarray}
\alpha_1\phi_1|_{x=0}=\alpha_2\phi_2|_{x=0}+\alpha_3\phi_3|_{x=0},\nonumber\\
\frac{1}{\alpha_1}\chi_1|_{x=0}=\frac{1}{\alpha_2}\chi_2|_{x=0}=\frac{1}{\alpha_3}\chi_3|_{x=0},\label{vbc01}
\end{eqnarray}
where $\alpha_1, \alpha_2, \alpha_3$ are the real constants which
will be determined below. In the following we will use
Eqs.~\re{vbc01} as the vertex boundary conditions for
Eq.~\re{nlde1} on a metric star graph. When applied to the exact
soliton solutions of the NLDE, the VBC will lead to algebraic
conditions connecting $\alpha_j$ and $g_j$, as will be derived in
the next section. We note that in the linear limit ($g_j\to 0$)
the vertex boundary conditions given by Eq.~\re{vbc01} preserve
the self-adjointness of the (linear) Dirac equation on metric
graph, i.e. belong to the class of general boundary conditions
derived by Bolte and Harrison in \ci{Bolte}.

\section{Soliton dynamics and vertex transmission}

In  \ci{Saxena10,Saxena12,Saxena14} the soliton solutions of the
nonlinear Dirac equation on the line have been obtained in the
form of standing wave solutions asymptoting to zero as $x\to \pm
\infty$. Here, we use the same definition for soliton solutions of
the nonlinear Dirac equation on the metric star graph presented in
Fig.~\ref{pic1}. We note in passing that for soliton solutions of
the NLDE derived in the Refs.\ci{Saxena10,Saxena12,Saxena14} all
elements of the energy-momentum tensor, except $T_{00}$  become
zero, i.e., $T_{01}=T_{10}=T_{11} =0$.

We look for the soliton solution of NLDE on a metric star graph in the form
\begin{equation}
\Psi_j(x,t)= \left(\begin{array}{cc}\psi_{1j}(x,t)\\
\psi_{2j}(x,t)\end{array}\right) = e^{-i\omega
t}\left(\begin{array}{cc}A_j(x)\\iB_j(x)\end{array}\right).\label{solit2}
\end{equation}
Then, from Eqs. \re{nlde1} and \re{solit2}  we have
\be
\frac{dA_j}{dx} + (m+\omega)B_j -g_j^2(A_j^2-B_j^2)B_j =0,
\lab{direq1}
\ee
\be
\frac{dB_j}{dx} + (m-\omega)A_j -g_j^2(A_j^2-B_j^2)A_j =0.
\lab{direq2}
\ee

The vertex boundary conditions for the functions $A_j$ and $B_j$ can be written as
\begin{eqnarray}
\alpha_1A_1|_{x=0}=\alpha_2A_2|_{x=0}+\alpha_3A_3|_{x=0},\nonumber\\
\frac{1}{\alpha_1}B_1|_{x=0}=\frac{1}{\alpha_2}B_2|_{x=0}=\frac{1}{\alpha_3}B_3|_{x=0}.\label{vbc004}
\end{eqnarray}

\begin{figure}[ht!]
 \includegraphics[width=140mm]{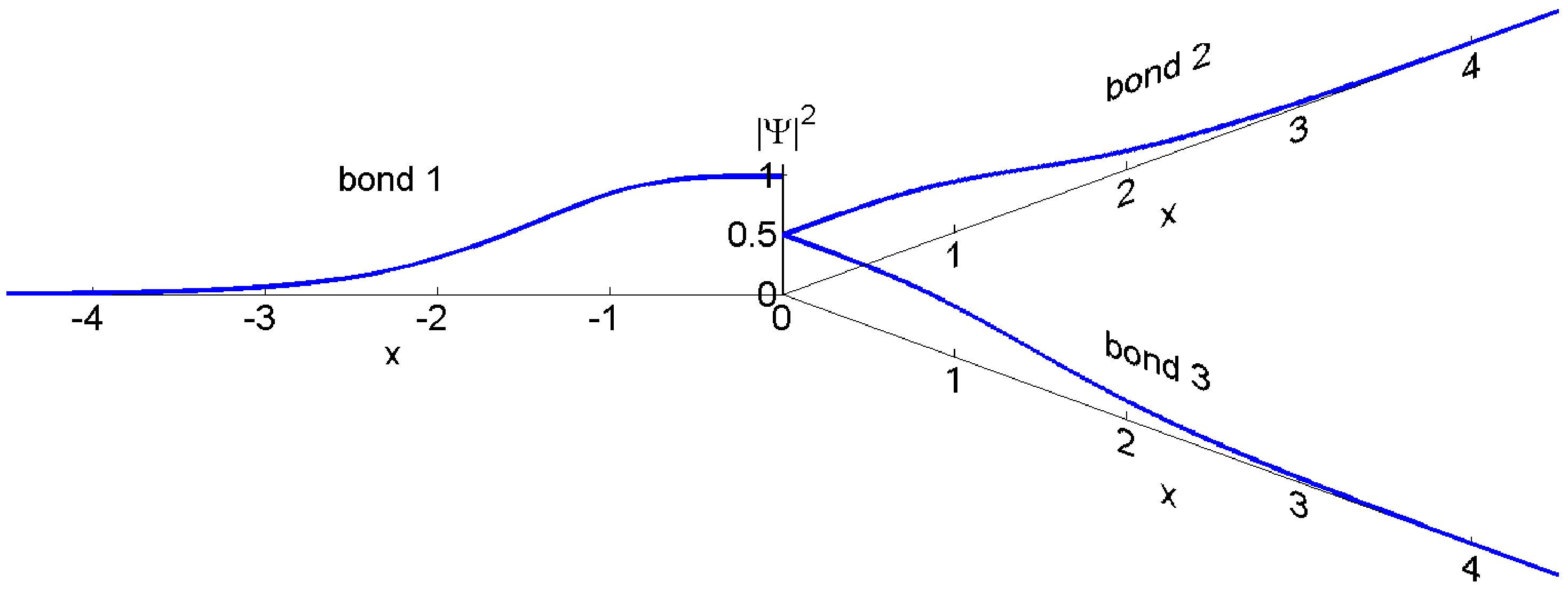} \caption{{(Color online) Charge density, $|\Psi_j|^2=\Psi^\dagger\Psi=|A_j(x)|^2+|B_j(x)|^2$,
  plotted using the standing wave soliton solutions given by  Eqs.~\re{solu1}
 and~\re{solu2}($x_0=0$) for $g_1=1,\;g_2=g_3=\sqrt{2}.$
\label{pic2}}}
\end{figure}

A prototypical static solution of the system \re{direq1},
\re{direq2} vanishing at $x \to \pm\infty$ can be written as
\ci{Saxena12} \baa A_j(x)=\sqrt{\frac{(m+\omega)\cosh^2(\beta
(x-x_0))}{m+\omega \cosh(2\beta (x-x_0))}}
\sqrt{\frac{2\beta^2}{g_j^2(m+\omega \cosh(2\beta (x-x_0)))}},
\lab{solu1} \eaa \baa B_j(x)=\sqrt{\frac{(m-\omega)\sinh^2(\beta
(x-x_0))}{m+\omega \cosh(2\beta
(x-x_0))}}\sqrt{\frac{2\beta^2}{g_j^2(m+\omega \cosh(2\beta
(x-x_0)))}}, \lab{solu2} \eaa where $x_0$ is the position of the
soliton's center of mass and $\beta=\sqrt{m^2-\omega^2}.$ In order
for these solutions of Eqs.~\re{direq1} and \re{direq2} to solve
the problem on the metric graph, they need to also satisfy the
vertex boundary conditions \re{vbc004}. This can be achieved if
the constants $\alpha_j$ and coupling parameters $g_j$ fulfill the
following conditions:
\begin{equation}
\frac{\alpha_1}{|g_1|}=\frac{\alpha_2}{|g_2|}+\frac{\alpha_3}{|g_3|}.\label{eq90}
\end{equation}
which stems from the first of Eqs.~(\ref{vbc01}) and
\begin{equation}
  \frac{\alpha_{2,3}}{\alpha_1}=\frac{|g_1|}{|g_{2,3}|},
\end{equation}
in accordance with the second one of conditions~(\ref{vbc01}).
The combination of the two leads to the ``sum rule'':
\begin{equation}
\frac{1}{g_1^2}=\frac{1}{g_2^2}+\frac{1}{g_3^2}.\label{eq9}
\end{equation}

It is important to note that this sum rule is derived by assuming
that incoming wave comes from the first bond, $b_1$, while if it
comes from the bond $b_2$ (or $b_3$) one should replace $g_1$ in
Eq.~\re{eq9}  with $g_2$ (or $g_3$). In Fig.~\ref{pic2} plots of
the soliton solutions of NLDE on metric star graph corresponding
to Eqs.~\re{solu1} and \re{solu2}, satisfying the vertex
conditions are given.

We note that  soliton solutions given by Eqs.\re{solu1} and
\re{solu2} describe the standing waves in metric graphs. The
traveling wave (soliton) solutions of NLDE can be obtained by
considering the case of the  moving frame and taking into account
the  Lorentz invariance \ci{Saxena12}, in the case of the
homogeneous domain (i.e., without a Y-junction).
  Assuming implicitly that the solitary wave is
  centered around a position well to the left of the origin, the Lorentz
transformations between the frames moving with relative velocity,
$v$ can be written as \ci{Saxena12}
\begin{equation}
x'=\gamma(x-vt);\; t'=\gamma(t-vx), \lab{LT}
\end{equation}
where \be \gamma =\frac{1}{\sqrt{1-v^2}}=\cosh\eta,\; \sinh\eta
=\frac{v}{\sqrt{1-v^2}}.\lab{LT1}\ee

Using these transformations, the traveling wave (soliton)
solutions of the nonlinear Dirac equation in the moving frame
determined by the constraints (\ref{eq90}) can be written as
\begin{eqnarray}
\psi_{1j}(x,t)&=&[\cosh(\eta/2)A_j(x')+i\sinh(\eta/2)B_j(x')]e^{-i\omega
t'},\nonumber\\
\Psi_{2j}(x,t)&=&[\sinh(\eta/2)A_j(x')+i\cosh(\eta/2)B_j(x')]e^{-i\omega
t'} \lab{wflt}
\end{eqnarray}
where $x',$ $t'$ and $\eta$ are given by Eqs.~\re{LT} and
\re{LT1}.

It is important to note that this transformation does not affect
the scaling dependence on the coupling parameters $g_j$. For that
reason, we expect the form of Eqs.~(\ref{eq90}) to still be valid
in the case of traveling/moving wave solutions.

\begin{figure}[t!]
\includegraphics[width=80mm]{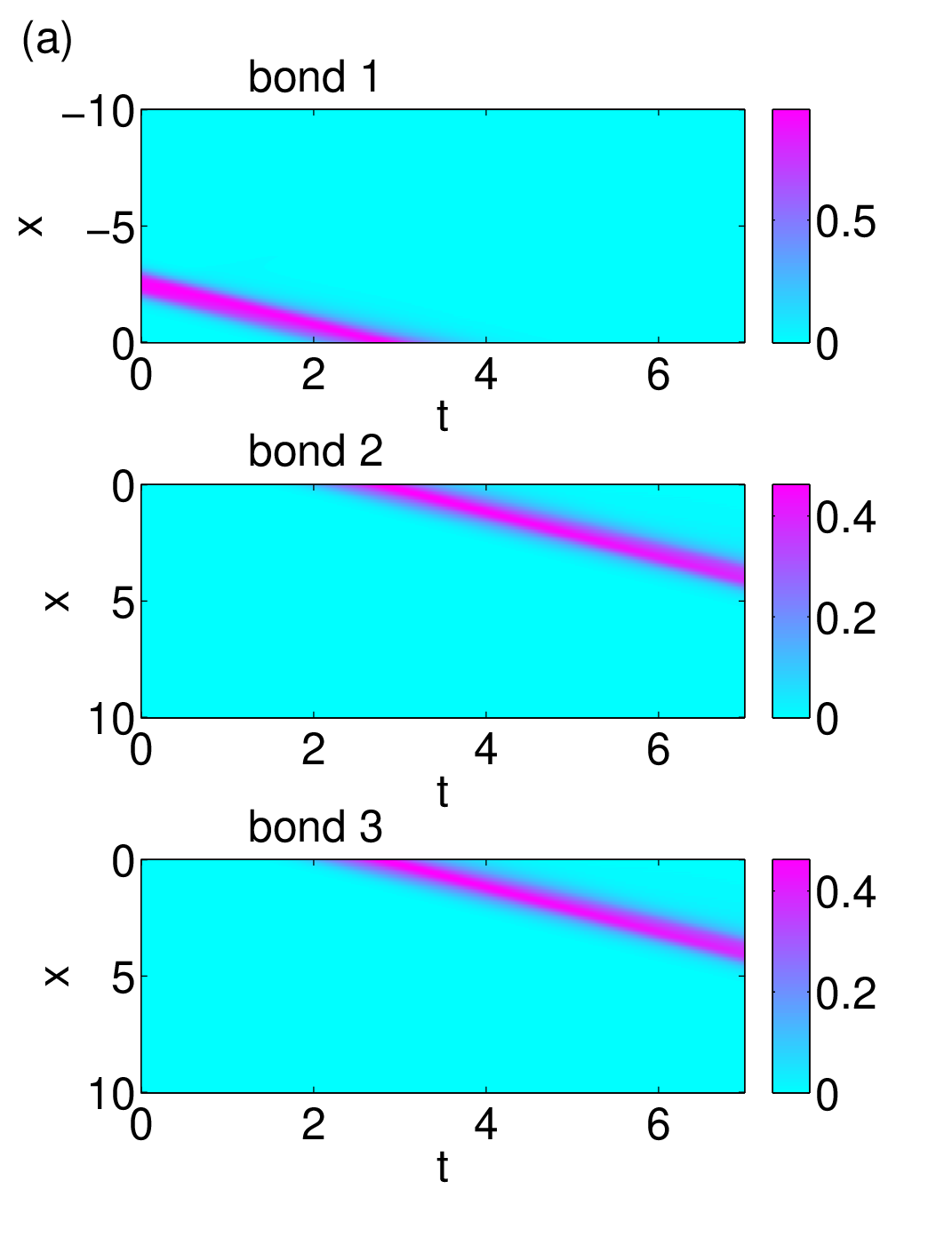}
\includegraphics[width=80mm]{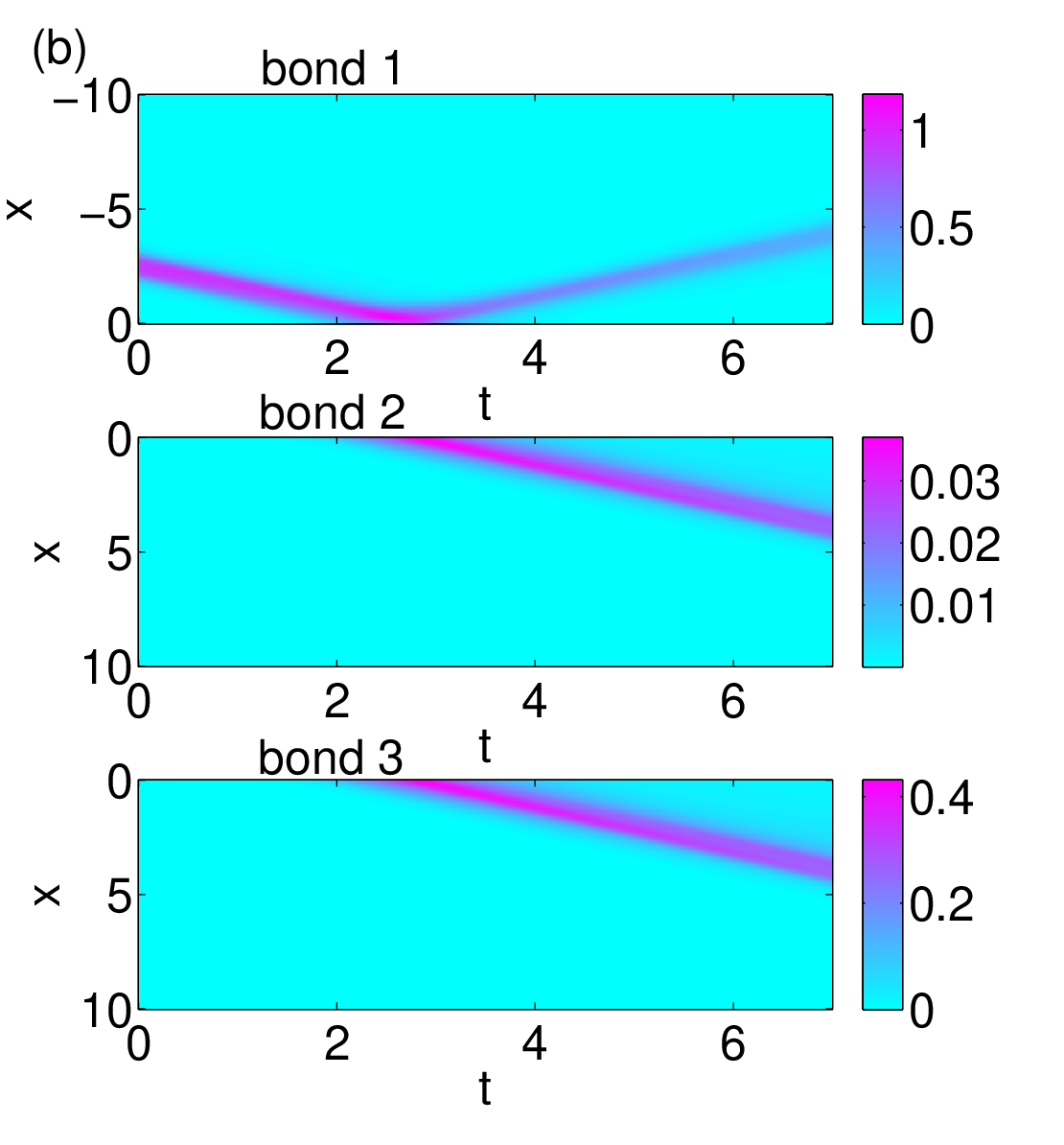}
\caption{{\small (Color online) Time and coordinate dependence of
the charge density obtained from the numerical solution of
Eq.~\re{nlde1} ($x_0=0$) for boundary conditions \re{vbc01} for
the case (a) when constraints \re{eq9} are fulfilled
($g_1=1,\;g_2=g_3=\sqrt{2}$), (b) when these constraints are not
fulfilled ($g_1=1,\;g_2=\sqrt{2}; g_3=3.5$). \label{pic3}}}
\end{figure}

\begin{figure}[t!]
\includegraphics[width=80mm]{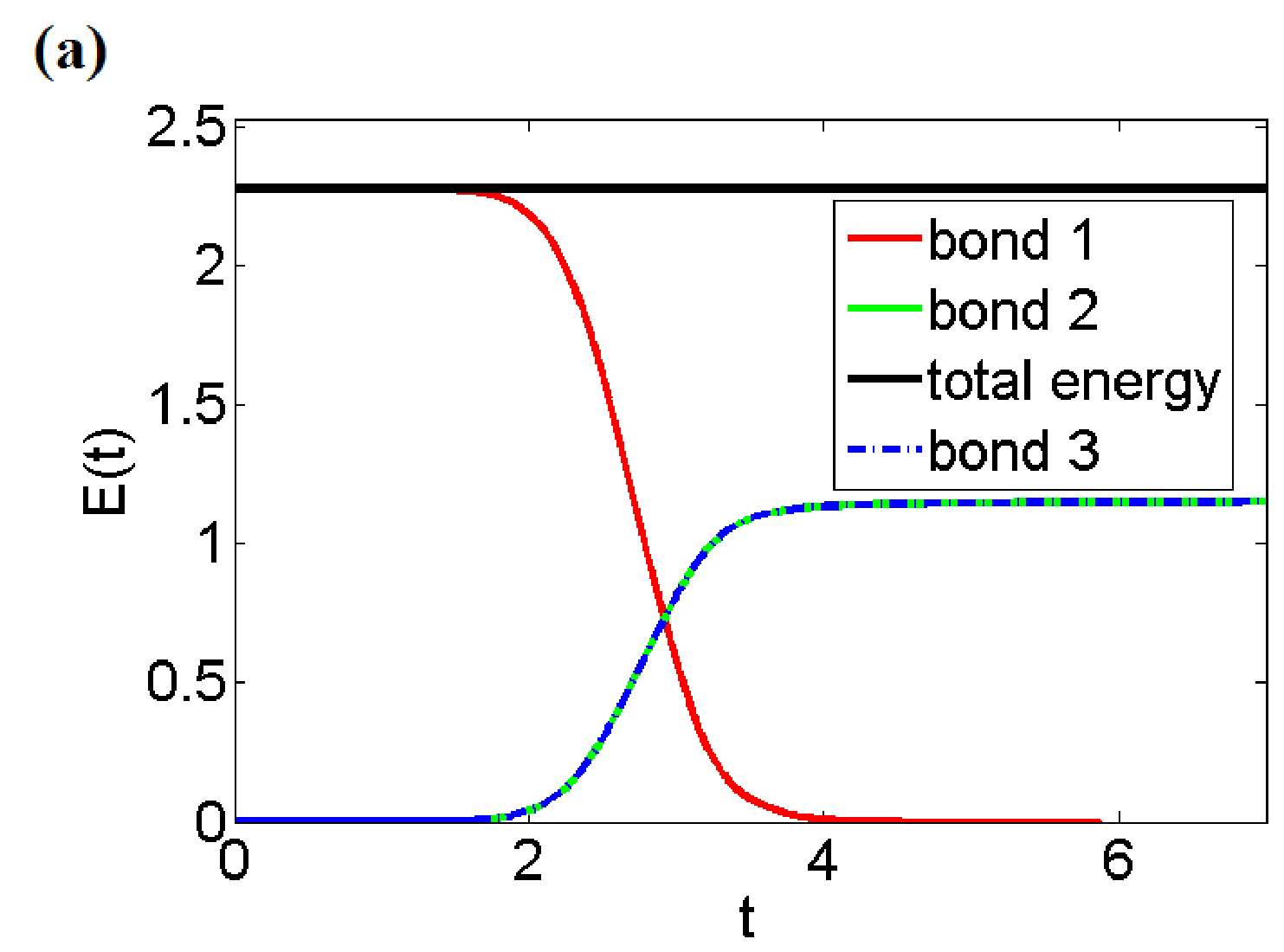}
\includegraphics[width=80mm]{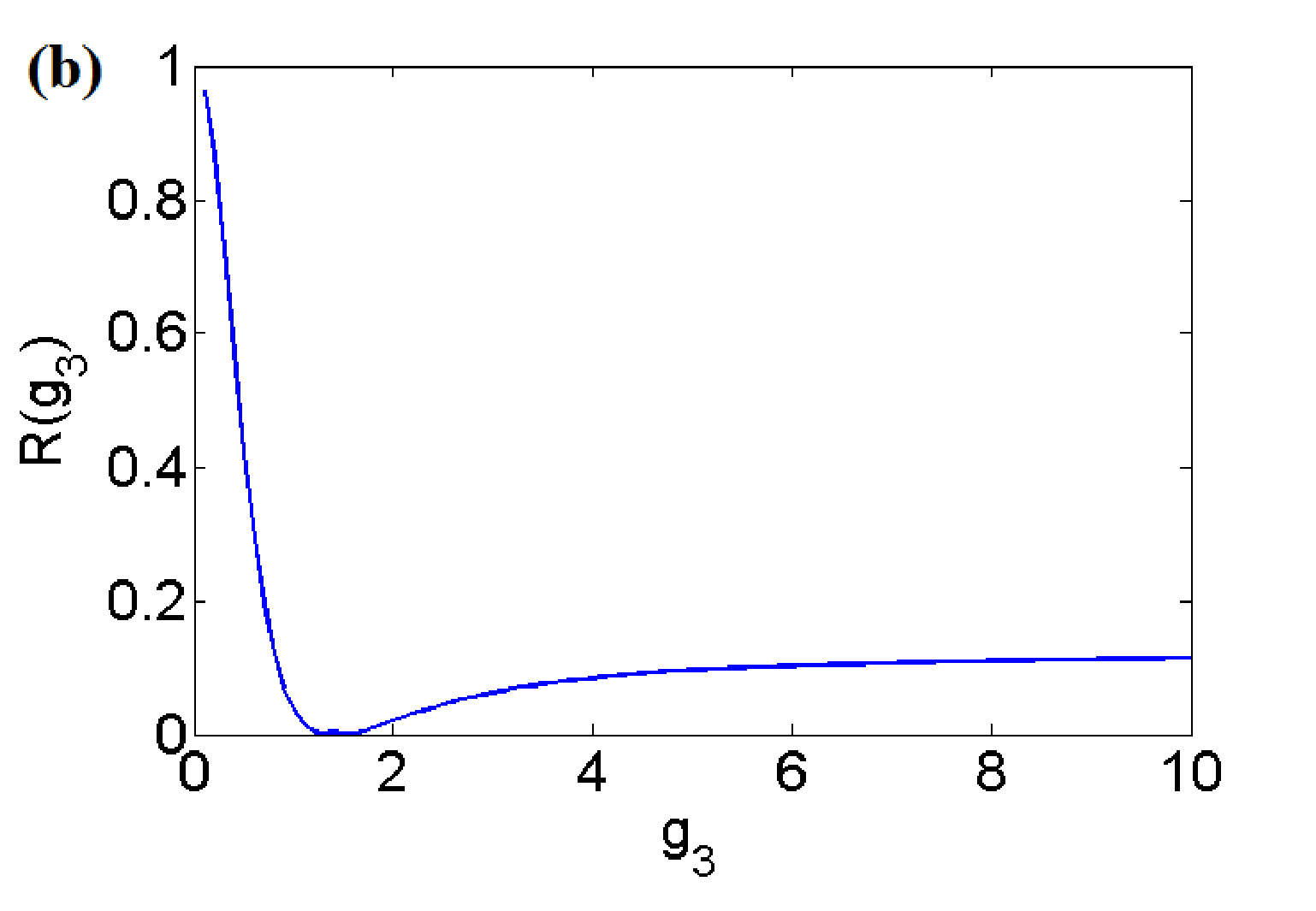}
\caption{(Color online) Numerical solution of Eq.\re{nlde1} for
boundary conditions \re{vbc01}. (a) Total energy and
time-dependence of the energy for each bond with
$g_1=1,\;g_2=g_3=\sqrt{2}$. (b) Reflection coefficient as a
function of $g_3$ for $g_1=1,\; g_2=\sqrt{2}$. } \label{pic4}
\end{figure}

To analyze the dynamics of the Dirac solitons on a metric graph,
we solve numerically the time-dependent NLDE given by
Eq.~\re{nlde1} for the vertex boundary conditions given by
Eq.~\re{vbc01} by considering the an initially traveling waveform
localized in bond 1, far from the vertex. We do this for the case
when the constraint given by Eq.~\re{eq9} is fulfilled, as well as
for the case of arbitrary $g_j$ which do not respect the relevant
constraints.
Fig. \ref{pic3}(a) presents the plots of the solution of
Eqs.~\re{nlde1} and \re{vbc01} obtained numerically for the values
of $g_j$ obeying the constraint \re{eq9}. For vertex boundary
conditions given by Eq.\re{vbc004} parameters $\alpha_j$ are
chosen as $\alpha_2=1/|g_2|$, $\alpha_3=1/|g_3|$,
$\alpha_1=|g_1|(\alpha_2/|g_2|+\alpha_3/|g_3|)$. One can observe
the absence of the vertex reflection in this case. Fig.
\ref{pic3}(b) presents soliton solutions of nonlinear Dirac
equation for the boundary conditions \re{vbc01} obtained
numerically for those values of $g_j$ which do not fulfill
Eq.~\re{eq9}. Here, reflection can be clearly discerned. Fig.
\ref{pic4}(a) shows the conservation of the energy and
reflectionless transmission through the Y-junction during the
propagation of Dirac soliton in graph. The emergence of the vertex
reflection when the constraints \re{eq9} are not fulfilled, can
also be seen from the Fig.~\ref{pic4}(b). Here the vertex
reflection coefficient, which is determined according to the
definition $R=E_1(t=7)/E(t=7)$ (with $E(t=7)
=E_1(t=7)+E_2(t=7)+E_3(t=7)$) is plotted as a function of $g_3$
for fixed values of $g_1$ and $g_2$. This systematic analysis
clearly illustrates the necessity of the symmetric scenario of
$g_3=\sqrt{2}$, consonant with our vertex sum rule, for reflection
to be absent. A similar phenomenon was observed in the case of
other nonlinear PDEs on metric graphs, such as
the NLS and sG equations, considered earlier in the
Refs.\ci{zar2010} and \ci{Our1}, respectively. For the initial
condition  corresponding to a traveling wave of the NLDE in the
form of \re{wflt} centered at $x_0$, with initial velocity $v > 0$
for $t \to + \infty$, our numerical computations provide strong
evidence to the following fact. The solution can be asymptotically
represented as  a superposition of solitary waves in the branches
associated with the second and third bonds, {\it only if} the
condition \re{vbc01} holds. In other words, under the necessary
constraints \re{eq9}, Eqs.~\re{vbc01} asymptotically lead to
refectionless transmission through the vertex.

\begin{figure}[htb]
\centerline{\includegraphics[width=90mm]{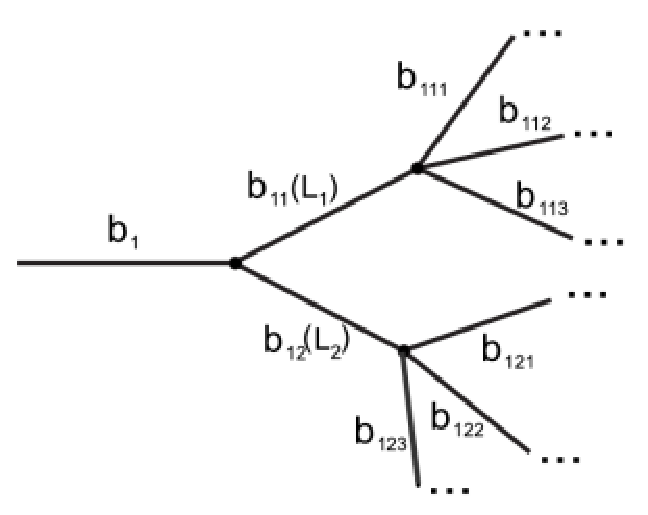}} \caption{Tree
graph. $x_1\in (-\infty, 0], x_{1i} \in [0,L_i]$, and $x_{1ij} \in
[0,+\infty)$ with $i=1,2;j=1,2,3$.} \label{tree-0}
\end{figure}

\section{Extension for other graphs}

The above treatment of the NLDE on the metric star graph can be
extended to the tree graph presented in Fig.~\ref{tree-0}. It
consists of three subgraphs $b_1, (b_{1i}), (b_{1ij})$, where
$i,j$ run over the given bonds of a subgraph. On each bond  $b_1,
b_{1i}, b_{1ij}$  we posit that the nonlinear Dirac equation  is
satisfied as given by Eq. \re{nlde1}. The vertex boundary
conditions can be written similarly to those in Eq.~\re{vbc004}.
Soliton solutions have a similar form to those in
Eqs.~\re{solu1}-\re{solu2} where $x$ is replaced by $x+s_b$, with
$s_1=s_{1i}=l,\,s_{1ij}=l+L_i$, with $l$ being position of the
center of mass of soliton. The sum rules
in this case generalize according to:
$$
\frac{1}{g_1^2}=\sum_{i=1}^2\frac{1}{g_{1i}^2},\,\,\,\frac{1}{g_{1i}^2}=\sum_{j=1}^3\frac{1}{g_{1ij}^2},\,\,\,i=1,2.
$$

\begin{figure}[htb]
\centerline{\includegraphics[width=90mm]{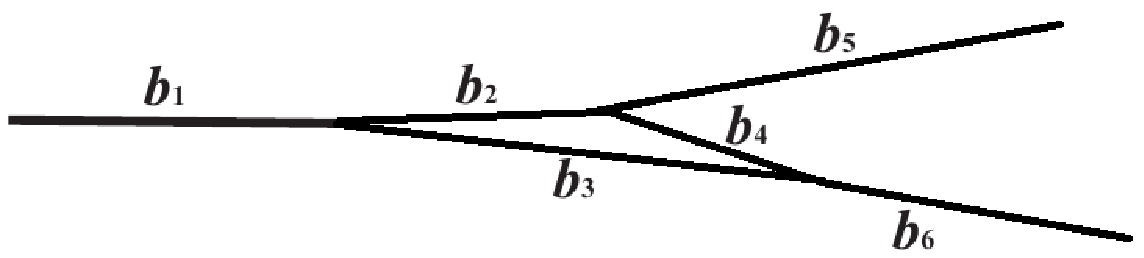}}
\caption{Triangle graph. $x_1\in (-\infty, 0], x_i \in [0,L_i]$
with $i=2,3,4$, and $x_{j} \in [0,+\infty)$ with $j=5,6$.}
\label{triangle}
\end{figure}

Similarly to that for star graph, one can obtain soliton solutions of
the nonlinear Dirac equation on the tree graph for the moving frame.
\begin{figure}[t!]
\includegraphics[width=85mm]{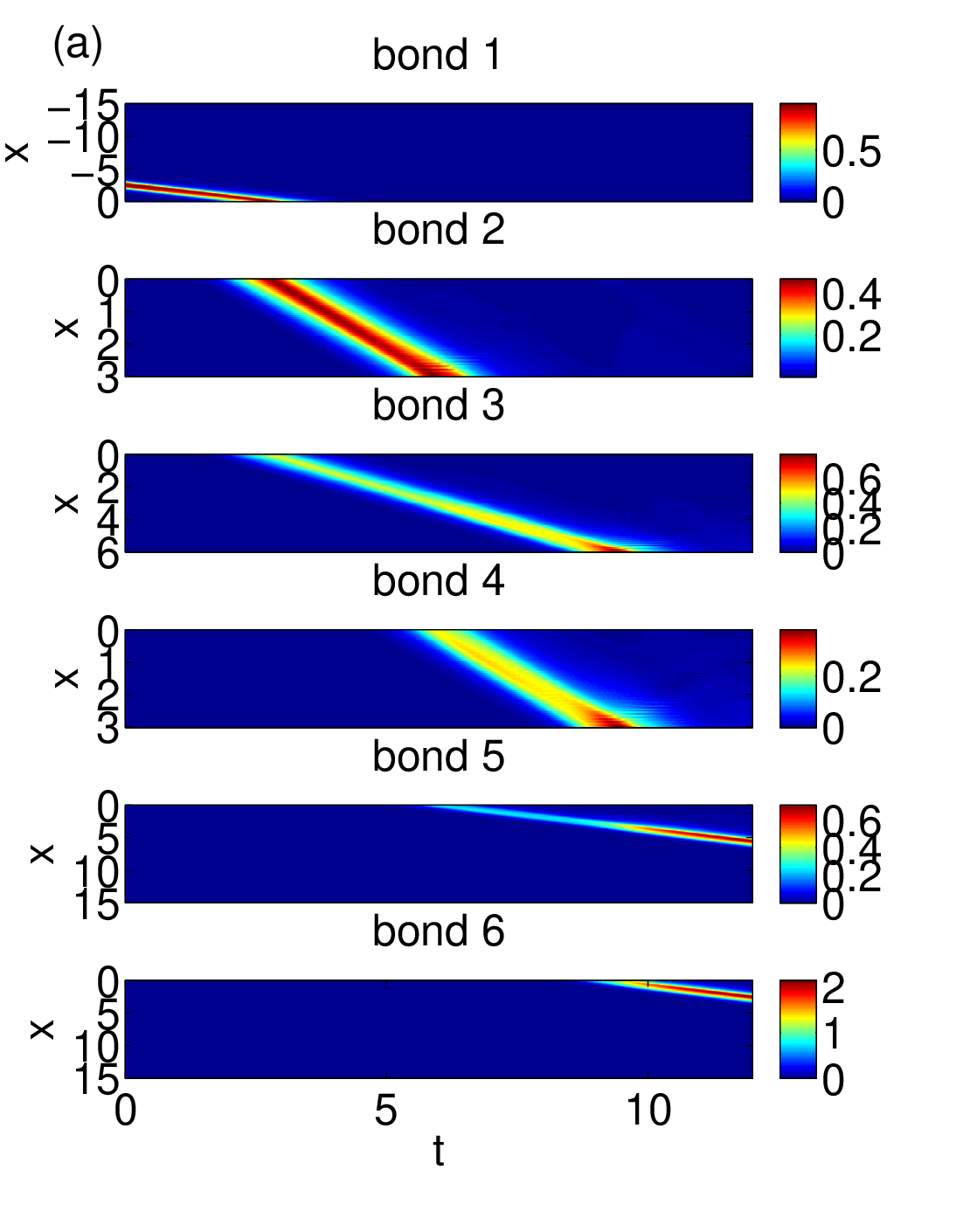}
\includegraphics[width=85mm]{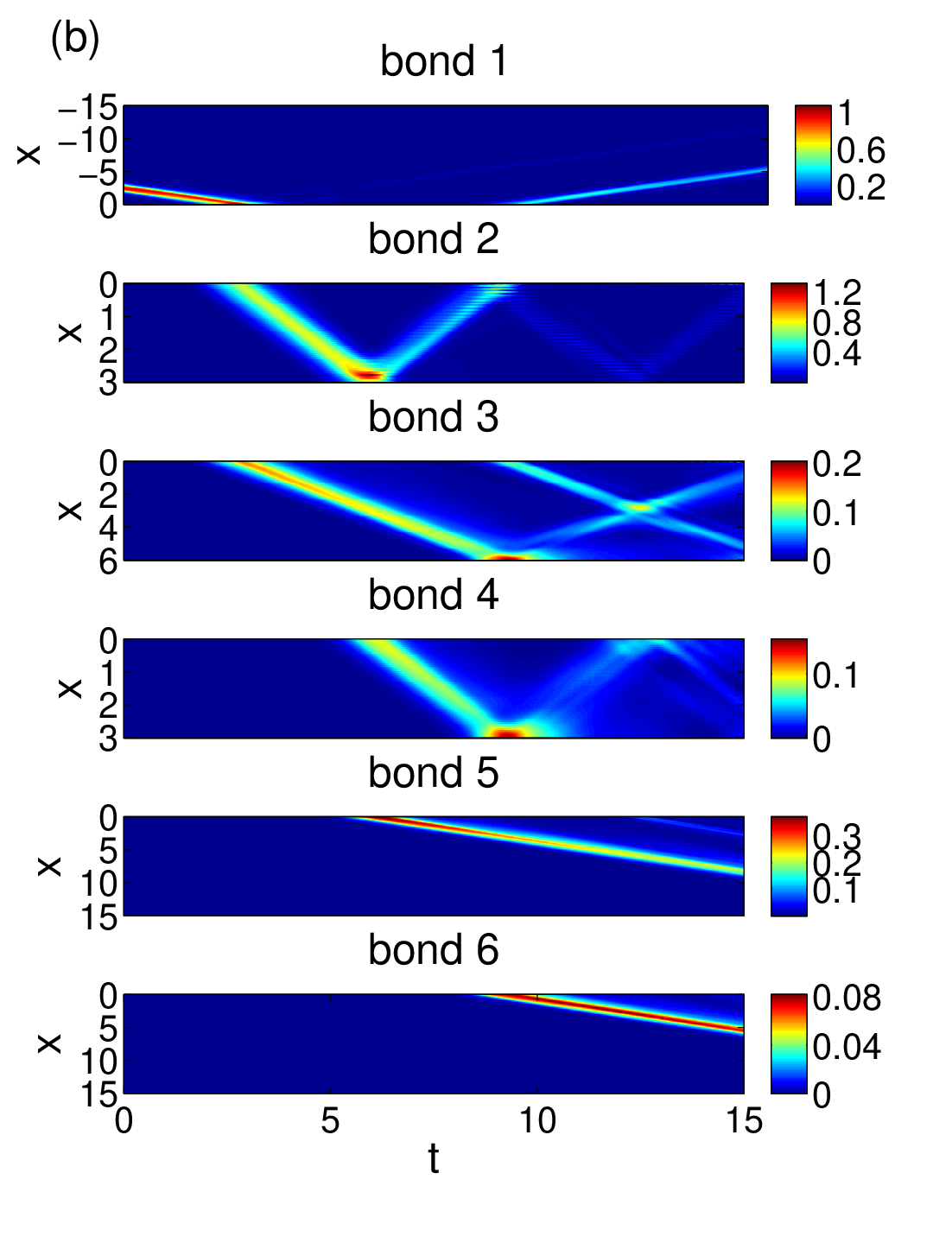}
\caption{{\small (Color online) Time and coordinate dependence of
the charge density obtained from the numerical solution of
Eq.\re{nlde1} ($x_0=0$)for the case
    (a) when constraints~(\ref{trianglecons1})--(\ref{trianglecons3})
    are fulfilled
($g_1=1,\;g_2=g_3=\sqrt{2},\;g_4=g_5=2,\;g_6=2/\sqrt{3}$), (b)
when these constraints are not fulfilled
($g_1=g_3=g_5=1,\;g_2=g_4=g_6=2$).  \label{pic3a}}}
\end{figure}

Another graph for which a soliton solution of the nonlinear Dirac
equation  can be obtained, is presented in  Fig.~\ref{triangle}. It
has the form of a triangle whose vertices are connected to outgoing
semi-infinite leads.
 The vertex boundary
 conditions for Eqs.~(\ref{direq1}) and (\ref{direq2}) which follow
 from the conservation of current and energy can be written as
\begin{eqnarray}
\alpha_1\phi_1|_{x=0}=\alpha_2\phi_2|_{x=0}+\alpha_3\phi_3|_{x=0},\nonumber\\
\frac{1}{\alpha_1}\chi_1|_{x=0}=\frac{1}{\alpha_2}\chi_2|_{x=0}=\frac{1}{\alpha_3}\chi_3|_{x=0},\nonumber\\
\alpha_2\phi_2|_{x=L_2}=\alpha_4\phi_4|_{x=0}+\alpha_5\phi_5|_{x=0},\nonumber\\
\frac{1}{\alpha_2}\chi_2|_{x=L_2}=\frac{1}{\alpha_4}\chi_4|_{x=0}=\frac{1}{\alpha_5}\chi_5|_{x=0},\nonumber\\
\alpha_3\phi_3|_{x=L_3}+\alpha_4\phi_4|_{x=L_4}=\alpha_6\phi_6|_{x=0},\nonumber\\
\frac{1}{\alpha_3}\chi_3|_{x=L_3}=\frac{1}{\alpha_4}\chi_4|_{x=L_4}=\frac{1}{\alpha_6}\chi_6|_{x=0}.\nonumber
\end{eqnarray}

In short, the conditions matching the of charge and energy
conservations must be applied to each node connecting the multiple
vertices of the graph. The soliton solution obeying these boundary
conditions can be written as
\begin{eqnarray}
A_b(x)=\frac{1}{|g_b|}\cosh(\beta
(x+s_b))\sqrt{\frac{m+\omega}{m+\omega
\cosh(2\beta (x+s_b))}}\nonumber\\
\times\sqrt{\frac{2\beta^2}{m+\omega \cosh(2\beta (x+s_b))}},\nonumber\\
\label{trianglesol1}
\end{eqnarray}
\begin{eqnarray}
B_b(x)=\frac{1}{|g_b|}|\sinh(\beta
(x+s_b))|\sqrt{\frac{m-\omega}{m+\omega
\cosh(2\beta (x+s_b))}}\nonumber\\
\times\sqrt{\frac{2\beta^2}{m+\omega \cosh(2\beta (x+s_b))}},\nonumber\\
\label{trianglesol2}
\end{eqnarray}
where $b$ is considered to be an index running from $1$ to $6$,
$s_1=s_2=s_3=l,\,s_4=l+L_2,\,s_5=l+L_2,\,s_6=l+L_3,\,L_3=L_2+L_4$
(this condition is necessary for the waves traveling along
$b_3$ and those split along $b_2$ and $b_4$ to arrive at
the vertex between $b_3$ and $b_4$ concurrently)
and
coefficients $g_j$ fulfill the following constraints:
\begin{eqnarray}
\frac{1}{g_1^2}=\frac{1}{g_2^2}+\frac{1}{g_3^2},\label{trianglecons1}\\
\frac{1}{g_2^2}=\frac{1}{g_4^2}+\frac{1}{g_5^2},\label{trianglecons2}\\
\frac{1}{g_3^2}+\frac{1}{g_4^2}=\frac{1}{g_6^2}.\label{trianglecons3}
\end{eqnarray}
Again, the traveling wave (soliton) solutions can be obtained in
case of the moving frame using the Lorentz transformations.
This approach for designing soliton solutions of the
nonlinear Dirac equation
can be applied to other graph topologies with multiple junctions,
provided a graph consists of finite parts connected to outgoing
semi-infinite bonds.

In Fig.~\ref{pic3a}  the solution of NLDE on each bond of this
triangle graph is plotted for the cases when the sum rule
given by Eqs.~(\ref{trianglecons1})-(\ref{trianglecons3})
is fulfilled (a)  and is broken (b).
Absence of the vertex reflection in Fig.~\ref{pic3a}(a) is clearly seen,
while  in Fig.~\ref{pic3a}(b) the transmission of Dirac solitons through the
graph vertices is clearly impeded by the mismatch at the VBC.
Thus, reflection events are clearly discernible in bonds such as
$1$, $3$ and $4$.

\section{Conclusions \& Future Challenges}
In this paper we studied dynamics of Dirac solitons in networks by
considering the case of metric graphs.
We obtained soliton solutions of the nonlinear Dirac equation on
some of the
simplest metric graphs such as the star, tree and triangle graphs.
Constraints enabling such exact solutions to exist are derived in the form of
sum rules for bond nonlinearity coefficients. The boundary conditions at the
branching points (vertices) are derived from
the fundamental conservation laws. It is shown
via direct numerical simulations that within
the corresponding dynamics the obtained constraints
provide for reflectionless transmission of solitary waves at the graph vertex.
In the case where the relevant conditions are violated, nontrivial
reflections ensue.

Our computations clearly suggest that the considered vertex
boundary conditions are necessary (yet not necessarily sufficient)
conditions for reflectionless transmission through the junctions
of interest. This is also in line with the recent homogenization
calculation of~\cite{dimarecent}. It is thus natural to formulate
the following conjecture. Assume that the initial data correspond
to a traveling wave of the NLDE in the form of Eq.~(\ref{wflt})
centered at $x_0$, with $v>0$. Then for $t \rightarrow +\infty$,
the solution can consist asymptotically of a superposition of
solitary waves in branches 2 and 3,  only if the condition of
Eq.~(\ref{eq9}) holds. This conjecture poses a challenging
mathematical question for further rigorous study in the problem at
hand, both at the level of the NLDE, but also by direct analogy in
the context of the NLS and other similar models.

It should be noted that an important, additional set of challenges
  emerges as regards the study of stationary solutions involving the
  vertex. In the language of the very recent work of~\cite{dimarecent},
  the states considered herein are the so-called half-soliton states.
  Yet in the NLS framework of the latter work, additional so-called
  shifted (non-monotone in the different branches) states also exist,
  yet they may be spectrally unstable, depending on whether they are
  monotonic or non-monotonic in the outgoing edges of the metric
  graph. The half-soliton considered here is especially interesting
  as at the NLS level it turns out to be spectrally stable, yet
  nonlinearly unstable. This issue of the (more involved at the
  NLDE level~\cite{comechh}) spectral stability of these stationary
  states is also an especially interesting direction for future
  study.

Naturally, the above study can be extended for other simple topologies,
such as a graph with one or multiple loops (e.g. the dumbbell graph), or
other
combinations of star, tree and loop graphs.

It will be extremely interesting if these ideas induce
experimental interest in systems such as suitably tailored
discrete optical waveguides~\ci{Conforti,Marini} or in atomic
settings, in the same way as they have for instance for the
propagation of traveling waves through networks of granular
crystals~\cite{daraio}.

\section*{Acknowledgement}
This work is partially supported by the grant of the Ministry for
Innovation Developments  of Uzbekistan (Ref. No.BF-2-022).

\section{Appendix A}
Here we will show that the linear vertex boundary conditions given by Eqs.~\re{vbc01}  lead to the ones given by~(\ref{vbc001})-(\ref{vbc002}).
Consider the following (linear) relations given at the vertex of the
a metric star graph
\begin{eqnarray}
\alpha_1\phi_1|_{x=0}=\alpha_2\phi_2|_{x=0}+\alpha_3\phi_3|_{x=0},\label{A1}\\
\frac{1}{\alpha_1}\chi_1|_{x=0}=\frac{1}{\alpha_2}\chi_2|_{x=0}=\frac{1}{\alpha_3}\chi_3|_{x=0},\label{A2}
\end{eqnarray}
where $\alpha_1,\,\alpha_2,\,\alpha_3$ are real constants.

Multiplying both sides
of Eq. (\ref{A1}) by $\frac{1}{\alpha_1}\chi_1^*|_{x=0},$ from Eq. (\ref{A2}) we
have
\begin{equation}
\phi_1\chi_1^*|_{x=0}=\phi_2\chi_2^*|_{x=0}+\phi_3\chi_3^*|_{x=0}.\label{A3}
\end{equation}
From Eq. (\ref{A3}) we get
\begin{equation}
{\rm Re}[\phi_1\chi_1^*]|_{x=0}={\rm Re}[\phi_2\chi_2^*]|_{x=0}+ {\rm Re}[\phi_3\chi_3^*]|_{x=0},\label{A4}
\end{equation}
which is nothing but the vertex boundary condition given by Eq.~\re{vbc001}.
Taking the time derivative from Eq. (\ref{A2}) and
multiplying  both sides of  (\ref{A1}) by $\frac{1}{\alpha_1}\partial_t\chi_1^*|_{x=0}$,  we obtain
\begin{equation}
\phi_1\partial_t\chi_1^*|_{x=0}=\phi_2\partial_t\chi_2^*|_{x=0}+\phi_3\partial_t\chi_3^*|_{x=0}.\label{A5}
\end{equation}
Taking the time-derivative from  Eq. (\ref{A1}) and multiplying the complex conjugate of the obtained expression by
$\frac{1}{\alpha_1}\chi_1|_{x=0}$ we get
\begin{equation}
\chi_1\partial_t\phi_1^*|_{x=0}=\chi_2\partial_t\phi_2^*|_{x=0}+\chi_3\partial_t\phi_3^*|_{x=0}.\label{A6}
\end{equation}
Adding Eqs. (\ref{A5}) and (\ref{A6})  leads to
\begin{eqnarray}
[\phi_1\partial_t\chi_1^*+\chi_1\partial_t\phi_1^*]|_{x=0}&=&[\phi_2\partial_t\chi_2^*+\chi_2\partial_t\phi_2^*]|_{x=0}\nonumber\\
&+&[\phi_3\partial_t\chi_3^*+\chi_3\partial_t\phi_3^*]|_{x=0}.\label{A7}
\end{eqnarray}
The last equation yields the vertex boundary condition given by Eq.~\re{vbc002}.


\begin{thebibliography}{99}

\bibitem{zar2010} Z. Sobirov, D. Matrasulov, K. Sabirov, S. Sawada, and K. Nakamura, Phys. Rev. E {\bf 81}, 066602 (2010).
\bibitem{zar2011} K. Nakamura, Z. A. Sobirov, D. U. Matrasulov, and S. Sawada, Phys. Rev. E {\bf 84}, 026609 (2011).
\bibitem{adami2011}  R. Adami, C. Cacciapuoti, D. Finco, and D. Noja, Rev. Math. Phys. {\bf 23}, 409 (2011).
\bibitem{adami-eur}  R. Adami, C. Cacciapuoti, D. Finco, and D. Noja, Europhys. Lett. {\bf 100}, 10003 (2012).
\bibitem{adami2013} R. Adami, D. Noja, and C. Ortoleva, J. Math. Phys. {\bf 54}, 013501 (2013).
\bibitem{Karim2013} K.K. Sabirov, Z.A. Sobirov, D. Babajanov, and D.U. Matrasulov,  Phys. Lett. A, {\bf 377}, 860 (2013).
\bibitem{noja}  D. Noja, Philos. Trans. R. Soc. A {\bf 372}, 20130002 (2014).
\bibitem{Susanto05}  G. H. Susanto and S.A. van Gils,  Phys. Lett. A 338, 239 (2005).
\bibitem{caputo14}  J.-G Caputo , D. Dutykh, Phys. Rev. E {\bf 90}, 022912 (2014).
\bibitem{Our} H. Uecker, D. Grieser, Z. Sobirov, D. Babajanov, and
D. Matrasulov, Phys. Rev. E {\bf 91}, 023209 (2015).
\bibitem{Our1} Z. Sobirov, D. Babajanov, D. Matrasulov, K. Nakamura,
and H. Uecker, EPL {\bf 115} , 50002 (2016).

\bibitem{Thirring} W. Thirring, Ann. Phys. {\bf 3}, 91 (1958).
\bibitem{Gross} D. J. Gross, A. Neveu, Phys. Rev. D {\bf 10}, 3235 (1974).
\bibitem{Steeb} W.H. Steeb, W. Oevel, and W. Strampp, J. Math. Phys. {\bf 25}, 2331 (1984).
\bibitem{Barashenkov1} I. V. Barashenkov and B. S. Getmanov, Commun. Math. Phys. {\bf 112}, 423 (1987).
\bibitem{Fushchich} W.I. Fushchych and R.Z. Zhdanov, J. Math. Phys. {\bf 32}, 3488 (1991).
\bibitem{Barashenkov2} I. V. Barashenkov and B. S. Getmanov, J. Math. Phys. {\bf 34}, 3054
(1992).
\bibitem{Toyama} F.M. Toyama, Y. Hosono, B. Ilyas and Y. Nogami,
  J. Phys. A: Math. Gen. {\bf 27}, 3139 (1994).
\bibitem{Ng} W.K. Ng, Int. J. Mod. Phys. A, {\bf 24}, 3476 (2009).
\bibitem{NG} W.K. Ng and R.R. Parwani, SIGMA {\bf 5}, 023 (2009).
\bibitem{Carr09} L. H. Haddad and L. D. Carr, Physica D {\bf 238}, 1413 (2009).
\bibitem{Carr11} L. H. Haddad and L. D. Carr, Europhys. Lett. {\bf 94} 56002 (2011).
\bibitem{Carr150} L.H. Haddad, C.M. Weaver and L.D Carr, New J. Phys. {\bf 17}, 063033 (2015).
\bibitem{Carr15} L.H. Haddad,  L.D Carr, New J. Phys. {\bf 17}, 063034
(2015).
\bibitem{Carr1501} L.H. Haddad,  L.D Carr, New J. Phys. {\bf 17}, 113011
(2015).
\bibitem{Carr151}  L.H. Haddad, K. M. O'Hara, L.D Carr, Phys. Rev. A  {\bf 91}, 043609 (2015).
\bibitem{Saxena10} F.Cooper, A. Khare, B. Mihaila and A. Saxena,  Phys. Rev. E  {\bf 82}, 036604 (2010).
\bibitem{Saxena12} F. G. Mertens, N. R. Quintero, F. Cooper, A. Khare, and
A. Saxena, Phys. Rev. E  {\bf 86}, 046602 (2012).
\bibitem{Pelinovsky12} D.E. Pelinovsky, A.Stefanov, J. Math.Phys.,{\bf 53}, 073705 (2012).
\bibitem{Pelinovsky14} D.E. Pelinovsky, Y. Shimabukuro, Lett. Mat. Phys.,{\bf 104}, 21 (2014).
\bibitem{Saxena14} S. Shao, F. G. Mertens, N. R. Quintero, F. Cooper, A. Khare, and A. Saxena, Phys. Rev. E  {\bf 90}, 032915 (2014).
\bibitem{Saxena15} J. Cuevas-Maraver, P.G Kevrekidis and A. Saxena,
  J. Phys. A: Math. Gen. {\bf 48}, 055204 (2015).
\bibitem{Saxena16} F. G. Mertens,  F. Cooper, N.R. Quintero, S. Shao, A. Khare, and A. Saxena, J. Phys. A: Math. Gen. {\bf 49}, 065402 (2016).
\bibitem{Saxena161} J. Cuevas-Maraver, P.G. Kevrekidis and A. Saxena, F. Cooper, A. Khare, A. Comech, C.M.Bender, IEEE J. Selected Topics Quant. Electr.,{\bf 22},  5000109  (2016).
\bibitem{Saxena162} J. Cuevas-Maraver, P.G Kevrekidis and A. Saxena,  A.Comech, R.Lan, Phys. Rev. Lett.  {\bf 116},  214101  (2016).
\bibitem{Pelinovsky16} A. Contreras, D.E. Pelinovsky, Y. Shimabukuro, Commun. Part. Diff. Eq.,{\bf 41}, 227 (2016).

\bibitem{Comech10} A. Komech, A. Komech, SIAM J. Math. Anal., {\bf 42},  2944 (2010).
\bibitem{Comech12} G. Berkolaiko, A. Comech, Math. Mod. of Natural Phenomena, {\bf 7}, 13  (2012).
\bibitem{Comech13} A. Comech, M. Guan and S. Gustafson, Ann. Henri Poincare (Analyse non lineaire), {\bf 31} 639 (2014).
\bibitem{Comech15} G. Berkolaiko, A. Comech, A. Sukhtayev, Nonlinearity, {\bf 28} 577 (2015).
\bibitem{Comech16} J. Cuevas-Maraver, P.G. Kevrekidis, A. Saxena, A. Comech, R. Lan, Phys. Rev. Lett. {\bf 116} 214101 (2016).

\bibitem{Zhang} Y Zhang, Nonlinear Analysis, {\bf 118}  82 (2015).
\bibitem{Lee} Ch-W. Lee,  P. Kurzynski,  H. Nha, Phys. Rev. A  {\bf 92} 052336  (2015).

\bibitem{Tran1} T.X. Tran,   X.N. Nguyen, F. Biancalana,
  Phys. Rev. A {\bf 91}, 023814 (2008).




\bibitem{Andriotis} A.N. Andriotis, M. Menon, Appl.Phys.Lett., {\bf 92} 042115 (2008).
\bibitem{Chen} Y.P. Chen, Y.E. Xie, L. Z. Sun and J. Zhong, Appl. Phys. Lett., {\bf 93} 092104 (2008).
\bibitem{Xu} J. G. Xu, L. Wang and M. Q. Weng, J. Appl. Phys., {\bf 114} 153701 (2013).
\bibitem{Kvashnin} A.G. Kvashnin, D.G. Kvashnin, O.P. Kvashnina and L.A. Chernozatonskii, Nanotechnology, {\bf 26} 385705 (2015).


\bibitem{Tran} T.X. Tran, S. Longhi, F. Biancalana, Ann. Phys.  {\bf 340}  179  (2014).

\bibitem{eugenieva} D.N. Christodoulides and E.D. Eugenieva Phys. Rev. Lett. {\bf 87}, 233901 (2001)

\bibitem{Bolte} J. Bolte and J. Harrison, J. Phys. A: Math. Gen. {\bf 36} 2747 (2003).
\bibitem{Conforti} M. Conforti, C. De Angelis, T. R. Akylas, and A.B. Aceves Phys. Rev. A {\bf 85} 063836  (2012).
\bibitem{Marini} A. Marini, S. Longhi, and F. Biancalana Phys. Rev. Lett. {\bf 113} 150401 (2014).

\bibitem{dimarecent} A. Kairzhan, D.E. Pelinovsky,  J. Phys. A: Math. Theor. {\bf 51}, 095203 (2018).

\bibitem{comechh} J. Cuevas-Maraver, N. Boussaid, A. Comech, R.Lan, P.G. Kevrekidis, A. Saxena,   arXiv:1707.01946.

\bibitem{daraio} A. Leonard, L. Ponson,  C. Daraio,  J. Mech. Phys. Solids {\bf 73}, 103 (2014).

\end{thebibliography}
\end{document}